\documentclass[twocolumn,preprintnumbers,amsmath,amssymb,floatfix]{revtex4}

\usepackage{graphicx}
\usepackage{color}
\usepackage{dcolumn}
\usepackage{amsmath,graphics,epsfig,color,verbatim}
\usepackage[linktocpage,bookmarksopen,bookmarksnumbered]{hyperref}
\newcommand{\vk}{{\mathbf{k}}}
 \newcommand{\vq}{{\mathbf{q}}}

\begin{document}

\setlength{\pdfpageheight}{\paperheight}
\setlength{\pdfpagewidth}{\paperwidth}


\title{Feynman's solution of the quintessential problem in solid state physics}


\author{Kun Chen}
\author{Kristjan Haule}
\email{haule@physics.rutgers.edu}

\affiliation{Department of Physics and Astronomy, Rutgers University, Piscataway, New Jersey 08854, USA}
\date{\today}





\maketitle

\textbf{ 
Two of the most influential ideas developed by Richard Feynman are
the Feynman diagram technique~\cite{feynman1949} and his variational
approach~\cite{feynman1955}.  The former provides a powerful tool to construct a systematic expansion for a generic interacting system,
while the latter allows optimization of a perturbation theory using a variational principle.
Here we show that combining a variational approach with a new
diagrammatic quantum Monte Carlo 
method~\cite{nikolay1998, prokof2008fermi,van2012feynman,VANHOUCKE201095, kozik2010diagrammatic,DMC_Hubbard,rossi2017, rossi2018}, both based on the
Feynman's original ideas, results in a powerful and accurate solver to
the generic solid state problem, in which a macroscopic number of
electrons interact by the long range Coulomb repulsion.
We apply the solver to the quintessential problem of solid state, the uniform
electron gas (UEG)~\cite{sommerfeld1928}, which is at the heart of the density
functional theory (DFT) success in describing real materials, yet it
has not been adequately solved for over 90 years. 
While some wave-function properties, like the ground state energy, have been very
accurately calculated by the diffusion Monte Carlo method (DMC)~\cite{ceperley1980}, the
static and dynamic response functions, which are directly accessed by
the experiment,
remain poorly understood.  Our method allows us to calculate the
momentum-frequency resolved spin response functions for the first
time, and to improve on the precision of the charge response function.
The accuracy of both response functions is sufficiently high, so as to
uncover previously missed fine structure in these responses.  This
method can be straightforwardly applied to a large number of
moderately interacting electron systems in the thermodynamic limit, including realistic models of metallic and semiconducting solids.}

The success of the Feynman's diagram technique rests on two pillars,
the quality of the chosen starting point, and one's ability to
compute the contributions of high-enough order, so that the sum
ultimately can be extrapolated to the infinite order. We will address
the former by introducing the variationally optimized starting point,
as discussed below, and we will solve the latter by developing a
powerful Monte Carlo method which can sum factorially large number of
diagrams while massively reducing the fermionic sign problem by
organizing Feynman diagrams into ``sign-blessed'' groups.

In the Feynman diagrammatic approach, one splits the Lagrangian of
a system, $L$, into a solvable part $L_0$ and the interaction
$\Delta L=L-L_0$. The effects of the interaction is included with a
power expansion in $\Delta L$, constructed using the Feynman
diagrams. Such diagrammatic series achieves the most rapid convergence
when the leading term $L_0$ captures the emergent collective behavior
of the system, which can be very different from the non-interacting
problem~\cite{anderson1972}. In the metallic state, which is of
special interest in this paper, the low temperature physics is
described by the emergent quasiparticles interacting with a screened
Coulomb interaction. We build an effective Feynman diagrammatic
approach by explicitly encoding such physics in $L_0$. We screen the
interaction in $L_0$ with a screening parameter $\lambda$,
rendering the Coulomb interaction short-ranged
($V(r) \propto exp(-r \sqrt{\lambda})/r$). Correspondingly, a
$\lambda$ counter-term must be added to $\Delta L$ to capture the
non-local effects of the Coulomb interaction with high order diagrams
(see the Methods section).  Similarly, we introduce an electron
potential $v_\vk$ which properly renormalizes the electron
dispersion and also fixes the Fermi surface of $L_0$ to the exact
physical volume, which is enforced by the Luttinger's
theorem~\cite{luttinger1960} (see the Methods section). In our
simulations, such choice shows the most rapid and uniform convergence of the response functions for both small and large momenta.


Motivated by Feynman's variational approach~\cite{feynman1955}, we
take the screening parameter $\lambda$ as variational parameters
which should be optimized to accelerate the rate of convergence. It
was shown in the development of optimized perturbation theory
~\cite{stevenson1981} and variational perturbation
theory~\cite{feynman1986,kleinert1995} that the best choice of a variational parameter is the value at which the targeted observable is least sensitive to the change of the parameter. This technique is called the principle of minimal sensitivity (PMS). In Refs.~\cite{stevenson1984,
  stevenson1985,stevenson1986,kleinert1995}, it was shown that the
perturbative expansion optimized with the PMS can succeed even when interaction is strong, and regular perturbation theory fails. In this work, we optimize the screening parameter $\lambda$ with PMS and observe a vast improvement to the convergence of the targeted response functions with expansion order.

 
%
\begin{figure}[bht!]
\includegraphics[width=0.99\linewidth]{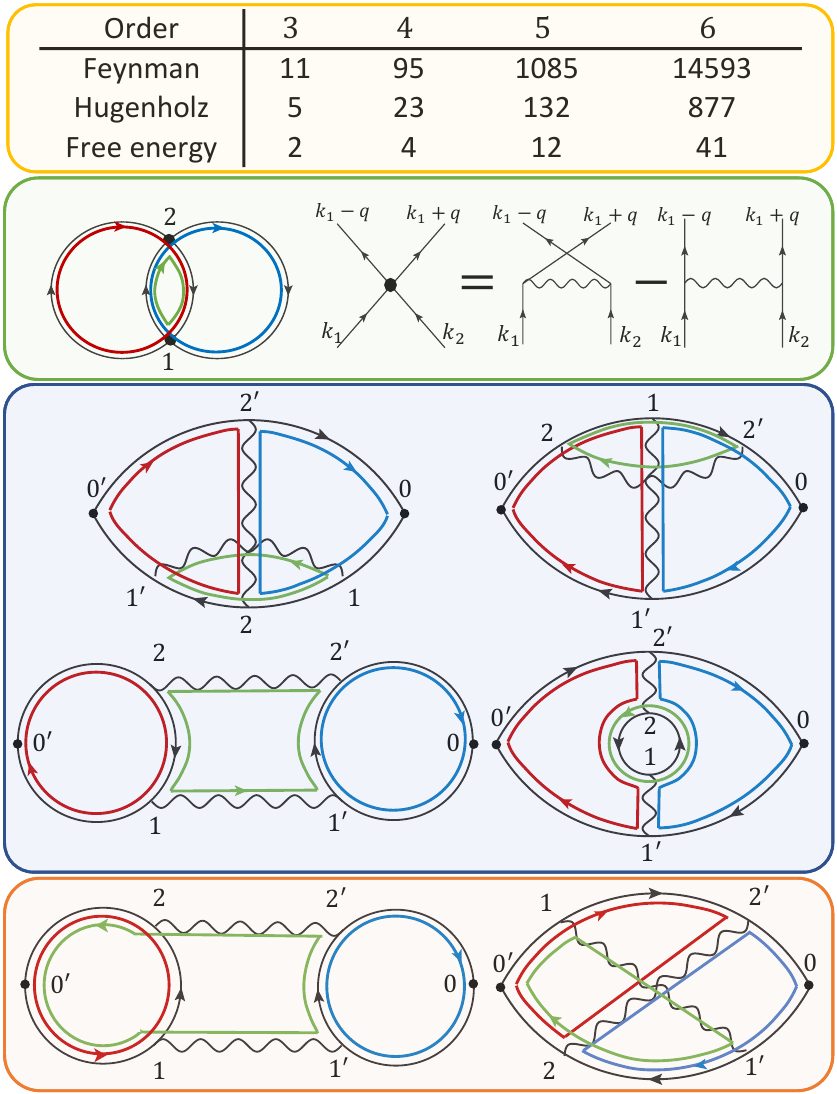}
\caption{ \textbf{The grouping of Feynman diagrams } is achieved by
  leveraging the fermionic crossing symmetry and the free-energy
  generating functional. Orange top box shows the number of
  Feynman/Hugenholtz/Free-energy Hugenholtz diagrams at orders 3-6, 
  excluding the Hartree-Fock sub-diagrams (see supplementary material).
  The green panel on the left and the right shows 
  an example of the free-energy Hugenholtz diagram, and how is the
  Hugenholtz vertex related to the standard Feynman diagram. Note that
  a single Hugenholtz diagram with $N$ vertices (black dots)
  represents up to $2^{N}$ standard Feynman diagrams with alternating
  signs.  
  By attaching two external vertices to different propagators in the
  Hugenholtz free energy diagram in the green box, one generates four
  topologically distinct groups of standard
  Feynman diagrams for the polarization function. Two of them are shown in the blue and orange box below.  
  By the process of attaching external vertices to a single
  Hugenholtz free energy diagram, we generate 10 out of 11 standard
  Feynman diagrams for the polarization at the third order.  The color
  lines represent our choice for momentum loops, which are uniquely determined by the choice of the loops in the free energy Hugenholtz diagram. 
  The external momentum is added through the shortest path connecting two external vertices.  
  Note that such grouping of diagrams allows us to calculate
  the weight of all diagrams in this figure with only 8 different
  electron propagators, instead of expected 36. 
  The above protocol can generate multiple copies of the same Feynman
  diagram (but with different choice of time and momenta), 
 which we weight with a proper symmetry factor}.
\label{Fig0}
\end{figure}

While our setup of the expansion (with the static screening and the
physical Fermi surface) is not entirely
new~\cite{kleinert1998,rossi2016,shankar1994,Counterterm}, its evaluation to high
enough order until ultimate convergence, has not been achieved before
in any realistic model containing long-range Coulomb interaction, as
relevant for realistic solids.  Our solution employs a recently
developed diagrammatic Monte Carlo
algorithm~\cite{nikolay1998,prokof2008fermi,van2012feynman,rossi2017, DMC_Hubbard}, which is
here optimized to take a maximal advantage of the sign blessing in
fermionic systems~\cite{prokof2008fermi}. Namely, by carefully arranging and grouping the Feynman diagrams, it is possible to ensure a massive sign cancellation for different diagrams in the same group, before the MC sampling is performed\cite{haule2010dynamical, rossi2017}.
 The previously used diagrammatic Monte Carlo algorithms,
which were sampling the diagrams one by one, are highly inefficient
here.


We evaluate diagrams in the momentum and imaginary-time representation, and for each configuration of random momenta
($\vk_0,\vk_1,\vk_2,\cdots,\vk_N$) and times
($\tau_1,\tau_2,\cdots,\tau_{2N}$) generated by the Markov chain, we sum
the contribution of all diagrams at a given order $N$, which have the
same number of momenta and time variables~\cite{haule2010dynamical}.
For example, when computing the polarization at order $N=6$, the sector
without counter-terms contains 14593 Feynman diagrams (see Fig.~\ref{Fig0}). 
These are regrouped into a much smaller number of
``sign-blessed'' groups to boost the efficiency of the MC sampling. 
For example, motivated by the crossing symmetry, at the lowest order in the
crossing exchange, we get from standard Feyman diagrams to so-called
Hugenholtz diagrams~\cite{hugenholtz1965} where the
direct and exchange interactions are combined into an antisymmetrized
four point vertex (see Fig.~\ref{Fig0} green box).
That exchange operation keeps the diagram exactly the same, except for a change
of the overall sign and a change of momentum on a single interaction line, hence the pairs of such diagrams largely cancel. After this operation,
there are only 877 Hugenholtz diagrams at order 6.
To further reduce the number of diagrams, we then combine the polarization diagrams that can be derived from the same free energy diagram by attaching two external vertices to propagators.
Mathematically, adding external vertices to a free energy diagram corresponds to taking its
functional derivative with respect to the inverse propagator. Therefore, the above step groups the polarization diagrams into a conserving group in the Baym-Kadanoff sense
~\cite{baym1961}, and the sign cancellation is guaranteed by the Ward
identities (See the supplementary material). 
For example, at order $N=6$ there are only
41 such free-energy groups (see Fig.~\ref{Fig0}).  We thus managed to
reduce the complexity from 14593 individual diagrams to 41 groups. 
The diagrams in the same group are very similar, and hence can share
the identical momentum/time variables (except the  external vertices). 
This not only ensures the massive sign cancellation between different
diagrams, but also reduces the cost of computing the total weight of Feynman diagrams
in Monte Carlo updates.


Finally, beyond variationally optimizing the zeroth order term ($L_0$)
we can also look for improvement of the high-order vertex
functions. One of our choices is to sum up all ladder diagrams
dressing the vertices (see the Method and Fig.3 in the supplementary material). We will call this scheme the Vertex Corrected Constant Fermi Surface (VCCFS).
The original diagrammatic expansion 
is here called Constant Fermi Surface (CFS)
scheme. The name originates in the above described principle that
electron potential $v_\vk$ is determined in such a way that $L$ and
$L_0$ share the same physical Fermi surface volume.

%
%
%
%
%

All results in this work are obtained at temperature $T=0.04 E_F$,
which is much lower than any other scale in the problem, hence results
are the zero temperature equivalent. We want to point out that finite
temperature calculations are very hard in the Diffusion Monte Carlo
(DMC), while our method is very well suited for finite temperature calculations, and converges even faster with the increasing order.
\begin{figure}[bht]
\includegraphics[width=1.0\linewidth]{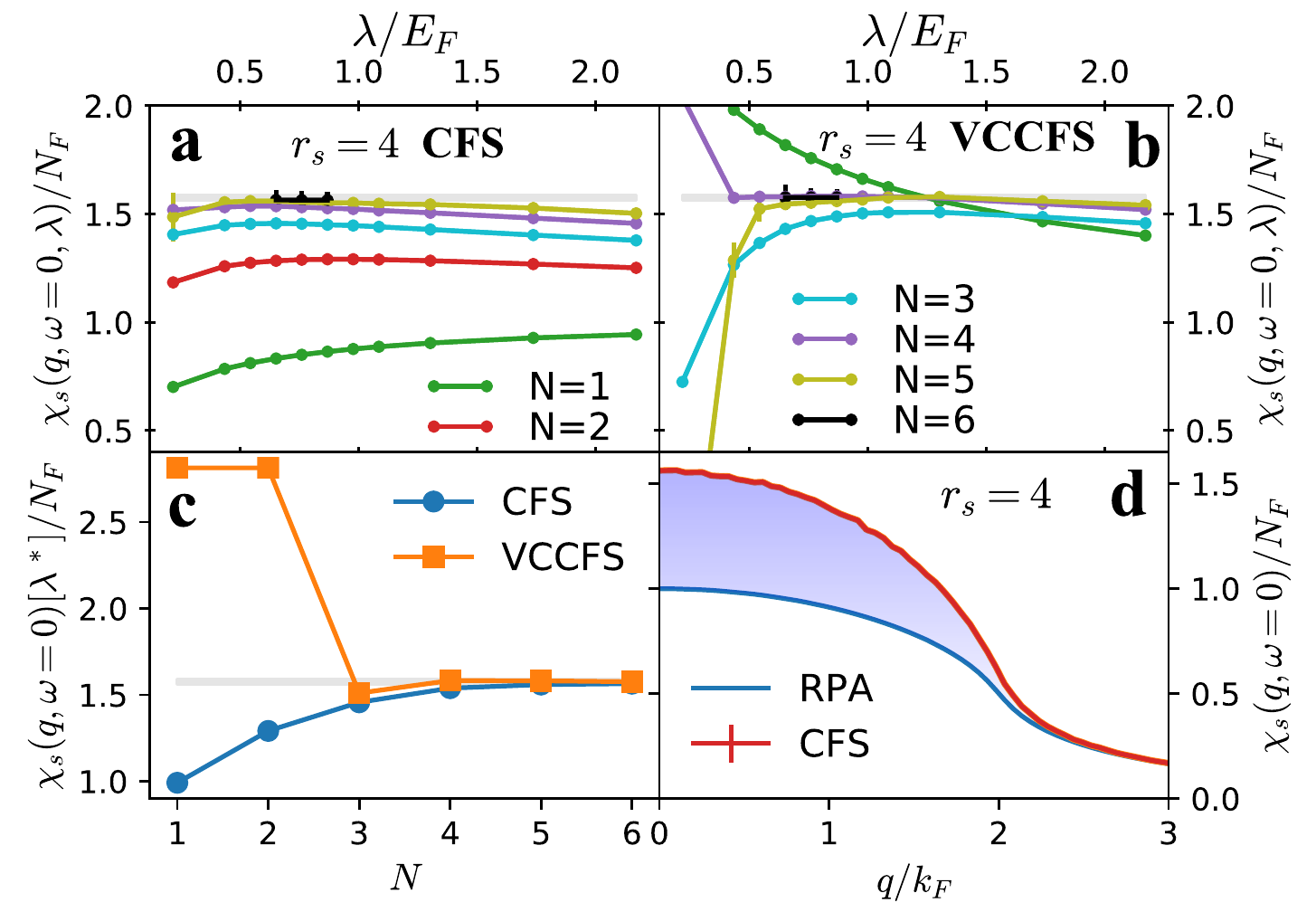}
\caption{
\textbf{Spin susceptibility of UEG at $r_s=4$}  (i.e., density $n=3/(4\pi r_s^3)$).
The optimization of $\chi_s(q=0,\omega=0)$ versus the screening parameter
$\lambda$ within (a) CFS  and (b) VCCFS scheme. Susceptibility $\chi$
and $\lambda$
are scaled by the density of states at the Fermi level
$N_F=(\frac{3}{2\pi})^{2/3}/(2\pi r_s)$, and the Fermi energy $E_F$,
respectively. The shaded region shows the estimated total error-bar of our calculation.
A single extrememum at the
optimized $\lambda^*$ appears, which is however order dependent ($\lambda^*_{N}$).
(c) The value of the optimized $\chi(q=0,\omega=0)[\lambda^*_{N}]$ versus
diagram order in both schemes. (d) The momentum dependent
$\chi(q,\omega=0)$ at the converged order $N=6$ and optimized
$\lambda^*_{N=6}/E_F=0.75$ in CFS scheme, along with 
comparison to Random phase approximation (RPA), which is exact when
interaction is ignored. The statistical errors are displayed in panels
(a), (b) and (d), and in (d) are smaller than the
width of the curve.
}
\label{Fig1}
\end{figure}
While wave-function properties, such as energy and pair distribution
function, are very precisely computed by 
DMC, and some of them are also are amenable to approximations such as
GW~\cite{van2017,kutepov2017one}, the response functions are more challenging to evaluate with the existing techniques. The strength of our approach is
that it can be used to compute both the static and the dynamic, the single and the multiparticle correlation functions. In
Figs.~\ref{Fig1} and \ref{Fig2} we show the momentum-dependent (Pauli) spin
susceptibility at zero frequency, which has never been precisely calculated  before to our knowledge even though its overall shape is crucial for the design of appropriate
exchange-correlation functionals of the DFT to predict magnetic order
in real materials. In panels (a) and (b) we show how the convergence properties of the 
susceptibility $\chi_s$ depends on the screening parameter $\lambda$ in the theory. 
Note that the static screening in $L_0$ is always compensated by the counter-term in $\Delta L$, 
so that for any value of $\lambda$ the UEG model is recovered at infinite order limit.
  The observable $\chi_s(q=0)$ develops a
broad plateau as a function of $\lambda$ (Fig.~\ref{Fig1}a and b) at the point
$\lambda^*_N$, which is slightly increasing with the increasing order.
This shows that if expansion is carried out to high enough order, the
physics becomes more and more local and allows one to use very short
range form of the interaction, which greatly improves the efficiency
of the method. We note that this property will be very beneficial in the realistic
material applications, where the converged result is extremely
difficult to obtain due to the long range nature of the bare Coulomb
interaction. Fig.~\ref{Fig1}c shows the value of $\chi_s(q=0)$ at the
optimized $\lambda^*_N$ versus perturbation orders. 
When the PMS is used, such that the variational parameter $\lambda$ is
optimized order by order, the convergence is very rapid, even when the
bare interaction is strong.  The value $\chi_s(q=0)$ at the optimized
$\lambda^*_N$ is monotonically increasing with the
increasing order in the CFS scheme, and beyond the second order is oscillating around the
converged value in VCCFS scheme. Both 
schemes converge towards the same value, and the
systematic error bar at a given truncation order can be estimated from
comparison between the two methods, 
allowing one to extract very
precise value of $\chi_s(q=0)$ even at a moderate expansion order (see
Fig.~\ref{Fig1}c and Table~\ref{Table2}).


Fig.~\ref{Fig1}d shows the momentum dependence of spin-susceptibility
$\chi_s(q)$ at $\lambda^*/E_F=0.75$, optimized at the highest order
($N=6$) and its comparison to the non-interacting (RPA) result, which
underestimates $\chi_s$ up to 57\%.

\begin{figure}[bht]
\includegraphics[width=1.0\linewidth]{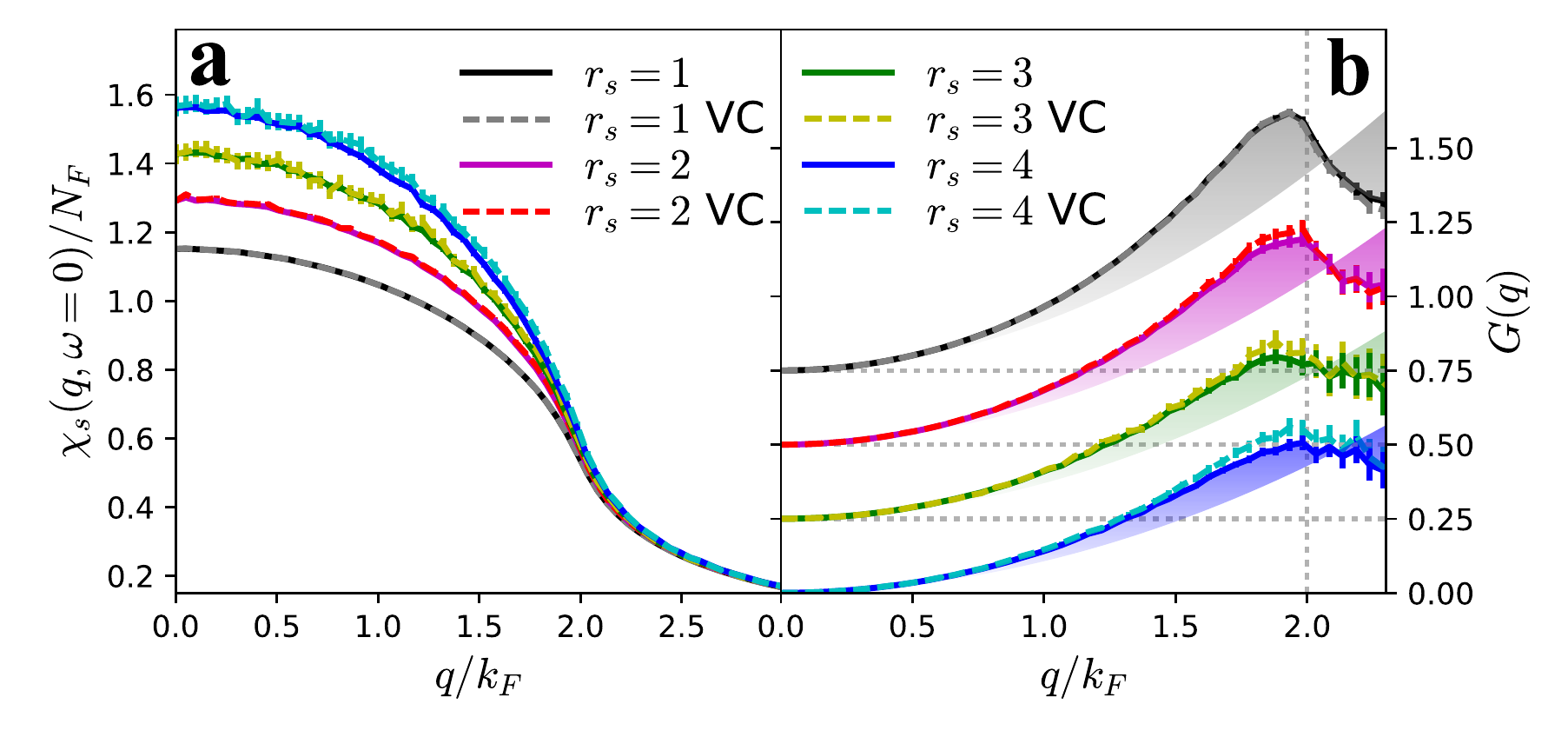}
\caption{
\textbf{Spin susceptibility:}
(a) $\chi_s(q,\omega=0)$ at optimized
$\lambda^*$ for $r_s=1-4$. 
VC corresponds to VCCFS scheme, and the rest to CFS scheme. The statistical error-bars are displayed for each computed point, 
and each point is computed statistically independently.
In VCCFS
scheme the statistical error-bars are larger than in CFS scheme, but
agree with each other within the error-bar.
(b) The local field
correction for the same $r_s=1-4$, and its deviation from quadratic
approximation (see the color envelope).
For clarity the curves for $r_s=1,2$ and $3$ are shifted up for $0.75$, $0.5$,
and $0.25$.
}
\label{Fig2}
\end{figure}
In Figs.~\ref{Fig2}a we show the same spin-susceptibility as in
Fig.~\ref{Fig1}d, but for other values of density parameter $r_s=1-4$
(here density $n=3/(4\pi r_s^3)$.).
Both VCCFS and CFS schemes agree with each other within the statistical
error-bar at order N=6 for all $r_s\le 4$.
We note that this spin susceptibility plays a central
role in construction of the DFT exchange-correlation kernel for
magnetically ordered systems.
Finally, Fig.~\ref{Fig2}b displays
the static local-field correction, which measures 
the deviation from the non-interacting electron gas ($\chi_{RPA}$), 
$G(q)\equiv
\frac{q^2}{8\pi}(\chi_{RPA}^{-1}(q,\omega=0)-\chi^{-1}(q,\omega=0))$. 
It
is a very sensitive measure of electron correlations. 
It has been suggested in
the literature that the possible peak near $k \sim 2k_F$ is of great
importance for understanding the quasiparticle
properties~\cite{simion2008}. Within
the local density approximation, the function $G(q)$ is approximated by the quadratic parabola depicted in Fig.~\ref{Fig2}b~\cite{moroni1995}, which is an excellent
approximation at small $q \le k_F$, but its deviation from the quadratic
function is very pronounced near $2k_F$. Note that within RPA $G(q)$
vanishes, as RPA does not take into account the exchange-correlation kernel.
We note that our calculation clearly shows that in the exact solution,
the local field correction
displays non-trivial maximum just above $2k_F$, which is obtained here for the first time.



\begin{figure}[bht]
\includegraphics[width=1.0\linewidth]{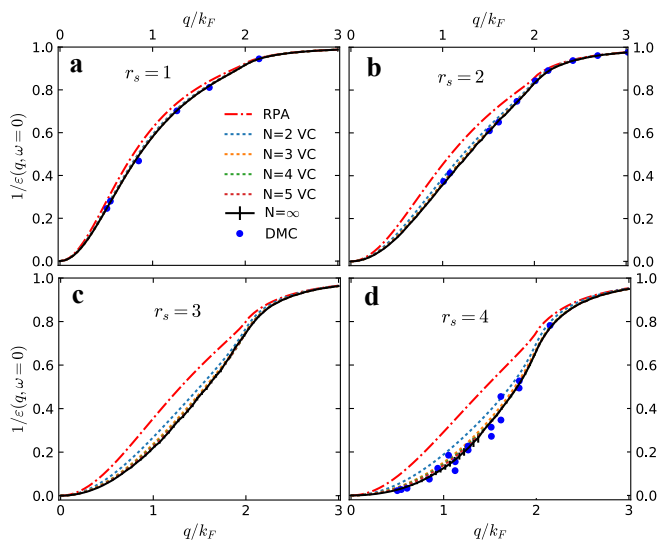}
\caption{\textbf{The inverse dielectric function }
($1/\epsilon$) for $r_s=1-4$
at $\lambda^*_{N=5}$, optimized for order 5, but we show
$1/\epsilon$ for all orders up to 5 and its extrapolated value. We
also display error-bars for extrapolated curve, which contains both
the statistical error, and the estimated extrapolation error.
Here we use more rapidly converging VCCFS scheme.
The comparison to DMC and RPA is shown. The DMC data are from
Ref. ~\cite{bowen1994, moroni1995}. 
}
\label{Fig3}
\end{figure}
Fig.~\ref{Fig3} shows the dielectric
function $\epsilon(q)$ for densities $r_s=1$ to $r_s=4$, and its comparison to RPA and
DMC~\cite{bowen1994, moroni1995} results. We show several orders ($N=2-5$) using
VCCFS scheme, and also the extrapolated result to $N=\infty$ using standard second order
Richardson extrapolation. 
The DMC data are in agreement
with our prediction, but notice that DMC allows one to calculate only
a set of discrete points, while the newly developed ``Variational Diagramatic Monte Carlo'' method gives
a smooth and very accurate continuous curve, which allows one to
resolve the fine structure.
For example, we notice that there is a clear kink
of $1/\epsilon$ curve near $2k_F$. This feature has been proposed in
some theories (e.g. Ref. ~\cite{utsumi1980}), but the previous
DMC results in Ref. ~\cite{bowen1994, moroni1995} were not precise
enough to confirm or disprove it.
%


%
%
\begin{table}[htb]
\begin{tabular}{l|ll|ll}
$r_s$ & $\chi_s/N_F$   & litt.($\chi_s/N_F$) & $P(0)/N_F$ & litt.($P(0)/N_F$)\\
\hline          
1     & 1.152(2) & 1.15-1.16               & 1.208(6)                      & 1.207-1.208  \\
2     & 1.296(6) & 1.27-1.31               & 1.54(2)                       & 1.549-1.549  \\
3     & 1.438(9) & 1.39-1.48               & 2.20(6)                       & 2.194-2.203  \\
4     & 1.576(9) & 1.51-1.66               &                                &                   
\end{tabular}
\caption{\textbf{Long wavelength values of spin and charge response:} 
First column $\chi_s=\chi_s(q=0,\omega=0)$ is the spin susceptibility, here normalized by the density of states at the Fermi level ($N_F$),
as computed by the current method. 
The second column shows the range of previous estimations from the literature~\cite{perdew1992accurate}.
$P(0)\equiv P(q=0,\omega=0)$ is the {static uniform} charge
polarization as obtained by this method. Unfortunately both CFS and
VCCFS methods approach the converged value from below, hence
extrapolation to $N=\infty$ is needed, which leads to much larger
error-bar in our calculation. 
The forth column lists previous DMC results, extracted from two
different correlation energy ansatzes proposed in 
Refs.~\onlinecite{perdew1992accurate} and ~\onlinecite{chachiyo2016communication}.
}
\label{Table2}
\end{table}
Finally, in Table~\ref{Table2} we give our best estimates for the
static spin and charge response with estimation of the error-bar. Within our method the spin response shows faster convergence with
increasing order, hence it allows us to compute the spin response more
precisely than the charge response, therefore our values for
$\chi_s/N_F$ are more precise than currently available literature (compare
columns one and two). Note that the previous estimate for the spin
susceptibility relied on an uncontrolled ansatz for the spin
dependence of the susceptibility, hence large uncertainty. 

Contrary to the spin response, or finite momentum charge response, 
the static uniform charge response
$P(q=0,\omega=0)$ can be obtained from the ground state energy of the
system, without explicitly introducing a modulated external potential,
and hence it can be extracted very precisely from the existing DMC calculations. We
compare it with our results, and find excellent
agreement. We note that static $P(q=0,\omega=0)$ at $r_s=4$
convergences very slowly in our method, due to proximity to the well
known charge instability at $r_s\approx 5.2$, hence we can not
reliably extrapolate its value to infinite order at $r_s\ge 4$.

The prospects of combining the Variational diagrammatic Monte Carlo
with DFT to obtain theoretically controlled results in real solids are
particularly exciting, as the DFT potential is semi-local and can be added
to $v_\vk$, so that it will play a role of a counter-term in the
expansion. The complexity would be modest, because no expensive
self-consistency is required, and because the interaction is statically screened even
at the lowest order, hence the scaling of this method should be
similar to the complexity of screened hybrids~\cite{ScreenedHybrids}
rather than the self-consistent GW approximation~\cite{RMartin}.

\textbf{Method}
The UEG model describes electrons in a solid where the
positive charges, which are the atomic nuclei, are assumed to be
uniformly distributed in space. The electrons interact with the other
charges through a long-range Coulomb interaction. The second-quantized
Hamiltonian is:
\begin{eqnarray}
\hat{H}=\sum_{\vk\sigma} \left( {\vk^2}-\mu \right) \hat{\psi}^\dagger_{\vk\sigma}\hat{\psi}_{\vk\sigma}+
\\
\frac{1}{2V}\sum_{\substack{\vq\ne 0\\ \vk\vk'\sigma\sigma'}}
\frac{8\pi}{q^2}\hat{\psi}^\dagger_{\vk+\vq\sigma}\hat{\psi}^\dagger_{\vk'-\vq\sigma'}\hat{\psi}_{\vk'\sigma'}\hat{\psi}_{\vk\sigma},
\end{eqnarray}
where $\hat{\psi}$/$\hat{\psi}^\dagger$ are the annihilation/creation operator of an electron, $\mu$ is the chemical potential controlling the density of the electron in the system.
We measure the energy in units of Rydbergs, and the wave number $k,q$ in units of inverse Bohr radius.

In the path integral representation, using the standard Hubbard-Stratonovich transformation, the Lagrangian
of the uniform electron gas can be cast into the form in which
the Coulomb interaction is mediated by an auxiliary bosonic field $\phi_\vq$. Motivated by the well known fact that the long-range Coulomb
interaction is screened in the solid, and that the effective potential
of emerging quasiparticles differs from the bare potential, we 
introduce the screening parameter $\lambda_\vq$ and an electron potential $v_\vk$ into $L_0$, which then takes the form
%
%
\begin{eqnarray}
L_0 = \sum_{\vk\sigma}\psi^\dagger_{\vk\sigma}
\left(\frac{\partial}{\partial\tau}-\mu+\vk^2+v_\vk(\xi=1)\right)\psi_{\vk\sigma}
\\
+\sum_{\vq\ne 0} \phi_{-\vq}
  \frac{q^2+\lambda_\vq}{8\pi}\phi_\vq,
\nonumber
\end{eqnarray}
and represents well the low-energy
degrees of freedom in the problem when parameters $\lambda_\vq$ and
$v_\vk$ are properly optimized. 
To compensate for this choice of $L_0$, we have to add the following interaction 
\begin{eqnarray}
\Delta L = 
-\sum_{\vk\sigma}\psi^\dagger_{\vk\sigma} v_\vk(\xi)\psi_{\vk\sigma}
-\xi \sum_{\vq\ne 0}\phi_{-\vq}
  \frac{\lambda_\vq}{8\pi}\phi_\vq
\\
+\sqrt{\xi}\frac{i}{\sqrt{2 V}}\sum_{\vq\ne 0} \left(\phi_{\vq}\rho_{-\vq}+\rho_{\vq}\phi_{-\vq}\right).
\label{Eq4}
\end{eqnarray}
so that, when the number $\xi$ is set to unity, $L = L_0+\Delta
L(\xi)$ is exactly the UEG Lagrangian.
The density $\rho$ is $\rho_\vq=\sum_{\vk\sigma}\psi^\dagger_{\vk\sigma}\psi_{\vk+q\sigma}$. 
Note that the first two terms in $\Delta L$ are the counterterms~\cite{Counterterm} 
which exactly cancel the two terms we added to $L_0$ above. 
We use the number $\xi$ to track the
order of the Feynman diagrams, so that order $N$ contribution sums up all
diagrams carrying the factor $\xi^N$. We set $\xi=1$ at
the end of the calculation.
Note also that this arrangement bears similarity with the well established
methods, such as G0W0~\cite{RMartin}, which computes the self-energy at the lowest
order ($\xi^1$) and sets $v_\vk$ to the DFT Kohn-Sham potential, and
$\lambda_\vq$ to the bubble diagram ($\lambda_\vq= g^0 g^0$ with ${g^0_\vk}^{-1}=(i\omega+\mu-\frac{\vk^2}{2m}-v_\vk)$).
The so-called skeleton Feynman diagram technique is recovered when $v_\vk$ and $\lambda_\vq$ are
equated with the self-consistently determined self-energy and polarization.
However, note that such diagram expansion can be dangerous, as it can lead to false convergence to
the wrong solution~\cite{French}

In optimizing the screening parameter $\lambda_\vq$ by the principle
of minimal sensitivity, we found it is sufficient to take a constant
$\lambda_\vq=\lambda$. Furthermore, we found that the uniform convergence
for all momenta is best achieved when the electron potential $v_\vk$
preserves the Fermi surface volume of $g^0_\vk$, therefore we expand
$v_\vk= \xi\,(\Sigma^x_\vk-\Sigma^x_{k_F}) + \xi^2\, s_2 + \xi^3\,
s_3\cdots$, and we determine $s_N$ so that all contributions at order
$N$ do not alter the physical volume of the Fermi surface. 
In other words, we ensure the density, which can be
  calculated with the identity $n=-P_{\vq}(\tau=0)$ where $|\vq| \gg
  k_F$, remains fixed order by order.
Since the
exchange ($\Sigma^x_\vk$) is static, and is typically large, we
accomodate it at the first order into the effective potential, so that
at the first order we recover the screened Hartree-Fock approximation,
i.e., interaction screened  to $\sim\exp(-r\sqrt{\lambda})/r$ and
optimized $\lambda$.

We also introduce a vertex correction scheme (VCCFS) to further improve the convergence of the series. In practice, within the VCCFS scheme, we
precompute the three-point ladder vertex, and attach it to both sides of a polarization Feynman
diagram, and at the same time, we eliminate
all ladder-type diagrams from the sampling, to avoid double-counting of diagrams (see the Supplementary Material).

Finally, we discuss the advantages and limitations of
  the proposed method.  The current variational approach is very
  effective at weak to intermeidate correlation strength (spin/charge
  response up to $r_s\approx 4$), but
  to extend it to the regime with stronger correlations, one would
  needs to introduce more sophisticated counter terms, such as the
 three and the four point vertex renormalization, to capture
  the emergent charge instability around $r_s\approx 5.2$.  Beyond the
  variational approach, we also want to point out that our developed
  Monte Carlo algorithm is a very generic Feynman diagram calculator
  for many-electron systems with long range Coulomb repulsion, and is
  more efficient and simpler that the existing conventional
  diagrammatic Monte Carlo of Refs
  ~\onlinecite{nikolay1998,prokof2008fermi,van2012feynman,VANHOUCKE201095,
    kozik2010diagrammatic,DMC_Hubbard}. For example, the new Monte
  Carlo algorithm requires only three updates, while the conventional
  approach needs about dozen updates. More importantly, this algorithm
  utilizes the ``sign-blessed'' grouping techinque to dramatically
  improve the sampling efficiency.
  Comparing to the recently proposed 
  Determinant Diagrammatic Monte Carlo algorithm~\cite{rossi2017}, our method is
  more generic in the sense that the algorithm can directly work in
  any representation (momentum/frequency, space/time) and can handle 
 any vertex renormalization
  withouth sacrificing the efficiency.

\textbf{Code Availability:} The code is available at
\url{https://github.com/haulek/VDMC}

\textbf{Acknowledgments:} 
We thank G.~Kotliar and N.~Prokof'ev and B. Svistunov and Y. Deng for stimulating discussion. This work is supported
by the Simons Collaboration on the Many Electron Problem, and NSF DMR--1709229.

\textbf{Author Contributions:} 
Both K.C. and K.H. developed the MC code, created the theoretical formalism, carried out the calculation, and
analyzed the results and wrote the paper. K.H. supervised the project.

\textbf{Competing financial interests}
The authors declare no competing financial interests.

\bibliographystyle{naturemag}
\bibliography{vDiagMC}{} 

\appendix
\section*{Supplementary Material}
\subsection{Conserving Diagrammatic Expansion}

This section introduces two conserving diagrammatic techniques, which
are called CFS and VCCFS in the main text, to calculate the
polarization $P$ (or susceptibility $\chi$). 
Both schemes preserve the exact crossing symmetry and conservation laws (particle number, momentum, energy, etc.) order by order.  
We note that the particle-number conservation law of the polarization $P({\bf{q}}\rightarrow 0, \tau)\rightarrow const$ is essential for the Coulomb electron gas, in order to properly describe 
the plasmon physics.  

The conserving diagrammatic expansions for the polarization can be constructed with the Baym-Kadanoff approach~\cite{baym1961,baym1962}, which is briefly reviewed below, before presenting the computational schemes used in the main text.
In the Baym-Kadanoff approach one first introduces an external potential coupled to the density operator of the system,
\begin{equation}
\label{eq:potential}
S[\psi^\dag, \psi; U]=S[\psi^\dag, \psi] - \int d1 d2 \; \psi^\dag (1) U(1,2) \psi(2),
\end{equation}
where $\psi$ are a Grassmann field for the electrons; the indexes represent spatial, temporal and spin variables. 
The generating functional for the connected correlation functions is defined as $\ln Z[U]$ with,
\begin{equation}
\label{eq:Z}
Z[U]=\int D\psi^\dag D\psi e^{-S[\psi^\dag, \psi; U]}.
\end{equation}
For a given approximation to $\ln Z[U]$, one can derive a conserving approximation for the one-particle Green's function by making sure that
\begin{equation}
\label{eq:green}
G(1,1')=\frac{\delta \ln Z[U]}{\delta U(1',1)}\bigg\vert_{U\rightarrow 0},
\end{equation}
while the two particle correlation function (charge, or spin correlation function if spin indexes are not summed), should satisfy
\begin{equation}
\label{eq:chi}
\chi(1,2)=\frac{\delta G(2,2^+; U)}{\delta U(1^+, 1)}\bigg\vert_{U\rightarrow 0},
\end{equation}
where the notation $1^+$ and $2^+$ indicates the time ordering of the field operators. 
The polarization, for which we will develop a diagrammatic expansion, is related to the correlation function $\chi$ by
%
\begin{equation}
\label{eq:polar}
\chi(1,2)=-P(1,2)+\int d3d3' P(1,3)v_{bare}(3,3')\chi(3',2),
\end{equation}
where $v_{bare}$ is the \textit{unscreened} Coulomb interaction. Note that the second term vanishes for the spin correlation function $\chi_{zz}$ in the unpolarized electron gas.

We will apply the above algorithm to the uniform electron gas model
defined by the Lagrangian $L=L_0+\Delta L$, where the solvable part is
\begin{eqnarray}
\label{eq:L0} 
L_0 = \sum_{\vk\sigma}\psi^\dagger_{\vk\sigma}
\left(\frac{\partial}{\partial\tau}-\mu+\vk^2+v_\vk(\xi=1)\right)\psi_{\vk\sigma}
\\
+\sum_{\vq\ne 0} \phi_{-\vq}
  \frac{q^2+\lambda_\vq}{8\pi}\phi_\vq,
\nonumber
\end{eqnarray}
%
and the correction is
\begin{eqnarray}
\label{eq:dL}
\Delta L = 
-\sum_{\vk\sigma}\psi^\dagger_{\vk\sigma} v_\vk(\xi)\psi_{\vk\sigma}
-\xi \sum_{\vq\ne 0}\phi_{-\vq}
  \frac{\lambda_\vq}{8\pi}\phi_\vq
\\
+\sqrt{\xi}\frac{i}{\sqrt{2 V}}\sum_{\vq\ne 0} \left(\phi_{\vq}\rho_{-\vq}+\rho_{\vq}\phi_{-\vq}\right).
\end{eqnarray}
%
This Lagrangian was introduced in the main part of the text.
Here the density $\rho$ is $\rho_\vq=\sum_{\vk\sigma}\psi^\dagger_{\vk\sigma}\psi_{\vk+q\sigma}$. Note that the effective potential $v_\vk$ and the inverse screening length $\lambda$ in $L_0$ are compensated by the counter-terms in the correction $\Delta L$. The parameter $\xi$ is set to unity at the end of the calculation. 

\begin{figure}[bht]
\includegraphics[width=0.9\linewidth]{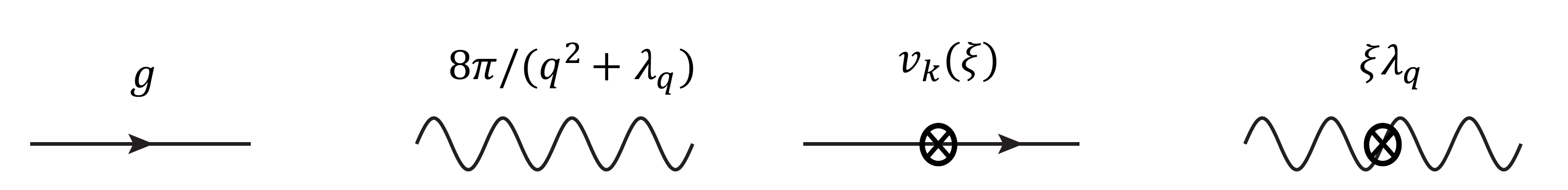}
\caption{
\textbf{Feynman diagram building blocks:}
The bare electron propagator $g$ describes an electron propagating in an effective potential $v_\vk$, 
the interaction line $8\pi/(q^2+\lambda_q)$ represents a bosonic propagator with an effective mass $\sim \lambda$, 
which makes the Coulomb repulsion short ranged. 
The counter terms, which compensate for our choice of the effective $L_0$, are proportional to 
$v_k(\xi)$ (the single-particle counter term) and $\xi \lambda_q$ (the interaction counter term), and are depicted in the last two diagrams.
}
\label{components}
\end{figure}

In the Baym-Kadanoff approach the external potential term $U(1,2)$ should be added to the solvable part $L_0$, 
and then the perturbative expansion for the generating functional $\ln Z[U]$ should be carried out using the standard Feynman diagrammatic expansion with building blocks
shown in Fig. \ref{components}. 
Note that the diagrammatic series constructed in this way only implicitly depends on the external potential $U$ through the bare electron propagator 
$g[U]^{-1}=-\frac{\partial}{\partial\tau}+\mu-\vk^2-v_\vk+U$. 

Now we are ready to discuss the Feynman diagrammatic expansion  used in our work. We will first discuss the CFS scheme.
To do this, we generate all free energy diagrams of order $N-1$, for example the diagram in Fig.1 of the main text, where the effective potential $v_\vk$ is regarded as an arbitrary function, independent of $U$. We then calculate the two-particle correlation function with the second derivatives with respect to external potential $U$,
$$\chi(1,2)=\left[\frac{\delta^2 \ln Z[U]}{\delta U(1^+,1)\delta U(2^+,2)}\right]_{v_\vk=\Sigma_\vk^x+\cdots, U=0},$$

Note that the $U$ derivative is taken by the chain rule, i.e., $\delta/\delta U=(\delta g/\delta U) (\delta/\delta g)$, where the the $U$-derivative of the propagator is simple: it just splits the propagator into two by inserting an external vertex,
\begin{equation}
\frac{\delta g(1,2; U)}{\delta U(3^+,3)}=-g(1,3)g(3,2).
\end{equation}
This relation is derived by taking the derivative of the identity $g^{-1} g=1$, which is $g^{-1} dg/dU+(dg^{-1}/dU) g=0$, therefore $dg/dU=-g (d g^{-1}/dU) g$ and $d g^{-1}/dU=1$, provided $v_\vk$ is independent of $U$. Diagrammatically, a derivative $\delta/\delta U$ removes a single-particle propagator from the Feynman diagram ($\delta/\delta g$), 
and we then replace it with an external vertex and the two propagators, i.e., $\delta g/\delta U=-g g$. 
In other words, it inserts an external vertex on an existing bare electron propagator. Note that this operation increases the diagram order by one. Finally, after the derivative is taken, we substitute $v_\vk$ with its expression in terms of the exchange self-energy,
\begin{equation}
v_\vk= \xi\,(\Sigma^x_\vk-\Sigma^x_{k_F}) + \xi^2\, s_2 + \xi^3\, s_3\cdots.
\label{Eq:vk}
\end{equation}

\begin{figure}[bht]
\includegraphics[width=1.0\linewidth]{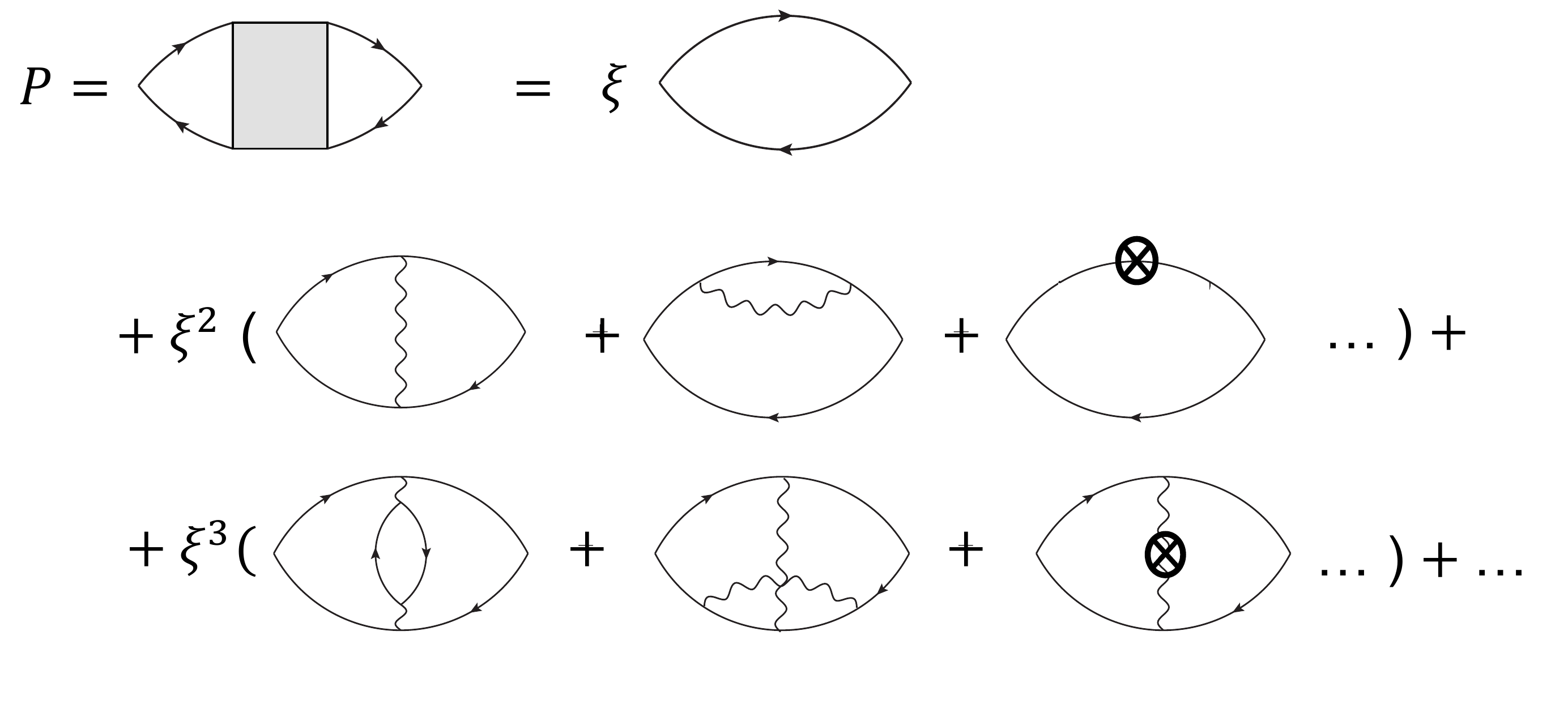}
\caption{
\textbf{Feynman diagrammatic expansion with counterterms:}
The perturbative expansion for the polarization is formulated with the standard Feynman diagrams with counterterms. 
The shaded block represents all one-interaction-irreducible diagrams for the particle-hole four-point vertex function. 
Note that the single-particle counterterm first appears at the second order, while the interaction counter-term first appears at the third order. 
Note that in this work we choose the single-particle counterterm to be the negative Fock diagram contribution plus a chemical potential shift, therefore
 any diagram with a Fock sub-diagram insertion (such as the diagram three above) is exactly canceled by a counter-term (such as the diagram four), and the two can hence be removed. 
 Consequently, one can simply drop the Fock sub-diagram insertion in the diagrammatic series, and keep only the chemical potential shift in the single-particle counter-term.
}
\label{diag}
\end{figure}

With the above described algorithm, we obtain the conserving expansion for the two particle correlation function $\chi$, however, 
the convergence for the dielectric function is even faster when the expansion is carried out for the polarization function defined by Eq.~\ref{eq:polar}. 
In the momentum and frequency space, the two are related by
\begin{eqnarray}
\chi(\vq) = -\frac{P_\vq}{1 - P_\vq \frac{8\pi}{\vq^2} }
\end{eqnarray}
or $\chi(\vq)=-[P_\vq+ P_\vq \frac{8\pi}{\vq^2} P_\vq + P_\vq \frac{8\pi}{\vq^2} P_\vq \frac{8\pi}{\vq^2} P_\vq+\cdots]$, meaning that $P_\vq$ is the irreducible part of $\chi(\vq)$ with respect to cutting the interaction propagator $\frac{8\pi}{\vq^2}$
 Similarly, when working with the screened interaction $\frac{8\pi}{\vq^2+\lambda}$, we can rewrite
\begin{eqnarray}
\frac{8\pi}{\vq^2}=\frac{8\pi}{\vq^2+\lambda}\sum_{n=0}^{\infty}\left(\frac{\xi \lambda}{8\pi} \frac{8\pi}{\vq^2+\lambda} \right)^n
\end{eqnarray}
and therefore
\begin{eqnarray}
\chi(\vq) = 
-\frac{P_\vq}{1 -P_\vq {\frac{8\pi}{\vq^2+\lambda}}{\sum_{n=0}^\infty \left(\frac{\xi\lambda}{8\pi} \frac{8\pi}{\vq^2+\lambda} \right)^n}},
\end{eqnarray}
which shows that $P_\vq$ is now the irreducible part of $\chi(\vq)$ with respect to cutting the interaction propagator $\frac{8\pi}{\vq^2+\lambda}$ or any
combination of interaction with counter terms of arbitrary order, i.e., $\frac{8\pi}{\vq^2+\lambda} (\frac{\xi\lambda}{8\pi}\frac{8\pi}{\vq^2+\lambda})^n $.
The resulting polarization diagrammatic expansion is shown in Fig. \ref{diag}. 

\begin{figure}[bht]
\includegraphics[width=1.0\linewidth]{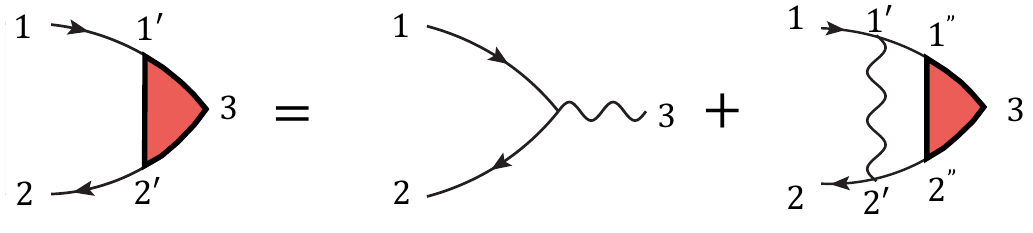}
\caption{
\textbf{Ladder-type vertex correction:}
{The ladder diagrams can be resummed by a Bethe-Salpeter equation.}
}
\label{diag_ladder}
\end{figure}

In practice, we find that the electric charge
  renormalization, which correspondes to the three-leg-vertex correction
  in diagrams, becomes increasingly more important at the low density
  limit (with $r_s \gtrsim 2$). 
Therefore, we introduce a vertex corrected scheme (VCCFS scheme),
where we resume all the ladder-type diagrams.

\begin{figure}[bht]
\includegraphics[width=0.9\linewidth]{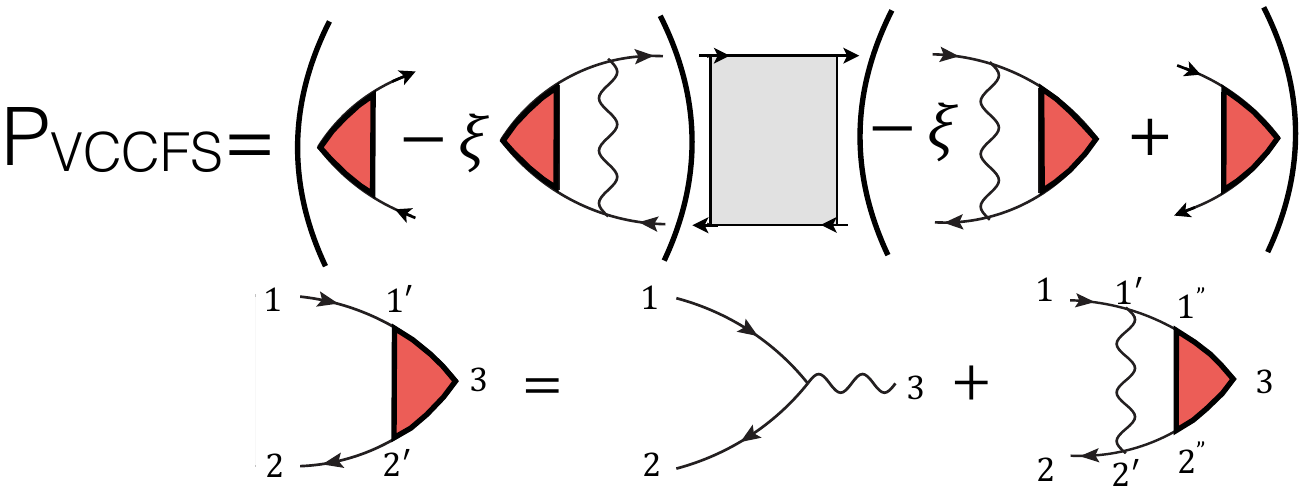}
\caption{
\textbf{VCCFS scheme for polarization diagrams:}
{The perturbative expansion for the polarization can be improved using the ladder resummation. 
The ladder vertex correction is attached to both sides (the left and right external vertex) and to all polarization diagrams. 
The double-counted diagrams are properly subtracted.}
}
\label{diag_polarization}
\end{figure}

The dressed ladder-type vertex correction can be
  calculated with a Bethe-Salpeter self-consistent equation, which is
  depicted in Fig.~\ref{diag_ladder}. In each polarization diagram, we
  then replace the two bare external vertices with the dressed
  vertices (the three-leg-vertex). 
To avoid the double counting of the diagrams, we also eliminate all
polarization diagrams which contains a ladder-type vertex correction
on either side of the diagram. 
This operation can be respresented by Fig. \ref{diag_polarization}, in
which the power expansion in powers of $\xi$ automatically removes all
diagrams with ladder-type vertex corrections on either end.

We emphasize here that all polarization diagrams in both schemes only involve the statically screened Coulomb interaction $\frac{8\pi}{\vq^2+\lambda_\vq}$.
This is a nontrivial result, given that the definition of the polarization in Eq. (\ref{eq:polar}) explicitly depends on the bare Coulomb interaction. 
Combing this feature with the fact that screened Coulomb interaction does not diverge in the long-wave-length limit, all polarization diagrams are now automatically regularized, making the Monte Carlo simulations much more efficient.



\subsection{Efficient Diagrammatic Monte Carlo Algorithm}
In this section, we introduce a simple yet efficient Monte Carlo
algorithm to evaluate high order Feynman diagrams. To calculate all order $N$ contributions, the diagrammatic Monte Carlo algorithm needs to integrate over all internal variables, such as momenta and times, and also sum over all topology of the diagrams, i.e.,
\begin{eqnarray}
F_1^N=\int [d\tau]^{2N} [d\vk]^{N+1} \sum_{topology} W[\{\tau\},\{\vk\}]
\end{eqnarray}
All diagrams in the same order share the same set of interval
variables. Due to the Fermi statistics, the sign of the integrand
$W[\{\tau\},\{\vk\}]$ alternates as the topology and internal
variables change. However, a Monte Carlo algorithm can only handle
positively defined weight functions. A straightforward choice is
to sample the absolute value of the integrand $|W[\{\tau\},\{\vk\}]|$, namely working with the sum,
\begin{eqnarray}
F_3^N=\int [d\tau]^{2N} [d\vk]^{N+1} \sum_{topology} |W[\{\tau\},\{\vk\}]|
\end{eqnarray}
However, as pointed out by the previous
studies~\cite{sign-blessing}, 
the sign
cancellation between diagrams causes $F_1^N \ll F_3^N$. More
specifically, although $F_3^N$ always diverge factorially with the
number of diagrams, the series $F_1^N$ is much better behaved
(diverging slowly, or even convergent if the series is within the
convergence radius). 
This phenomenon is termed the ``sign blessing" in Ref.~\cite{sign-blessing}. 
As a result, the straightforward Monte Carlo scheme sampling $F_3^N$ to evaluate $F_1^N$ suffers from the notorious sign problem, and is very inefficient.
In this work, we propose a Monte Carlo algorithm, which samples the following weight function,
\begin{eqnarray}
F_2^N=\int [d\tau]^{2N} [d\vk]^{N+1} |\sum_{topology} W[\{\tau\},\{\vk\}]|
\end{eqnarray}
Thanks to the inequality $F_1^N\le F_2^N \le F_3^N$, a method sampling
$F_2^N$ is guaranteed to suffer less sign problem, thus is more
efficient than the straightforward approach. 
Of course, the efficiency of this approach relies on how small is
$F_2^N$, and how close is $F_2^N$ to $F_1^N$.
The minimization of $F_2^N$ can be achieved by optimizing the
arrangement of interval variables of different diagrams, 
so that the sum of their weights with the same set of variables
strongly cancel with each other. 
We will discuss this in more detail in the next section.

Now we summarize the main steps of the new diagrammatic Monte Carlo
algorithm used in this work.
\begin{itemize}
\item[i)] Write a script to generate all Feynman diagrams up to the desired
truncation order (say order $6$ in this work), including all necessary
symmetry factors and counter-terms.
\item[ii)] Design an algorithm to properly assign interval variables to
minimize the weight function $F_N^2$ (choice of basis). This algorithm will be described in the next section.
\item[iii)] Use the standard Metropolis algorithm to sample$F_N^2$ in order to
calculate the high dimensional integral $F_N^1$. 
To properly normalize the integral $F_N^1$, we design an ansatz for a function, which can be
integrated deterministically, and has parameters that can be adapted
to the landsacpe of $F_N^2$.
\end{itemize}
Note that the Monte Carlo updates only need to randomly generate
internal variables $\vk$ and $\tau$, but do not need to change the
diagram topology, so that the algorithm is extremely simple. 

\subsection{``Sign-blessed" Group of Diagrams}
In this section we explain the details of our algorithm to organize
diagrams, so that an efficient diagrammatic Monte Carlo method can be
implemented.
We will show how the diagrams of a given order can be divided into groups, where the diagrams in the same group are guaranteed to massively cancel with each other. The ``sign-blessed" group may be obtained by grouping: i) diagrams that share the same set of internal variables, and those diagrams in which ii) the integrand $W[\{\tau\},\{\vk\}]$ massively compensate with each other. The first requirement is automatically satisfied for the connected diagrams of the same order $N$, as all order-$N$ connected diagram requires $N+1$ independent momentum/frequency variables or $2N$ space/time variables. The second requirement is much more challenging and can only be achieved by carefully examining the sign structure of the diagrams.

\begin{figure}[bht]
\includegraphics[width=0.7\linewidth]{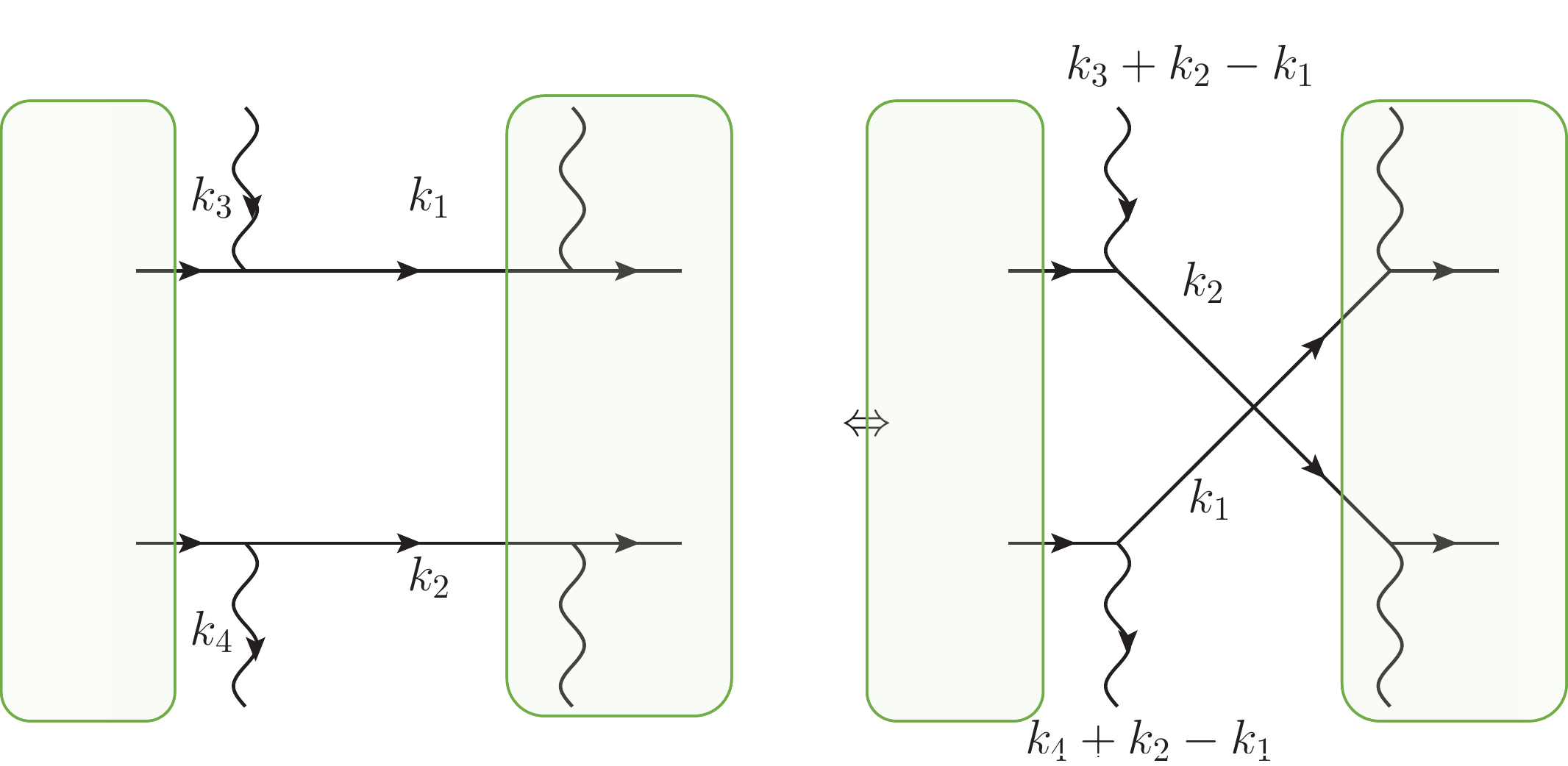}
\caption{ \textbf{Permutation using the crossing symmetry:}
  A permutation of a two fermionic propagators, as shown in the
  unshaded region of the figure, creates a new diagram with
  the opposite sign.  If the incoming  momentums/frequencies are
  similar,  the two diagrams have almost the same absolute value and opposite
  sign, hence the sum of the two diagrams leads to ``sign
  blessing''. Under this operation, the rest of the diagram, together with the shaded
  regions, remains the same. In order to achieve the cancellation of
  the integrand (not just the resulting integral), the two diagrams
  have to be consistently labeled, and therefore the
  momentum/frequency labeling of the entire diagram outside the unshaded region  has to be identical on the two diagrams.}
\label{bases}
\end{figure}

We identify two useful generic rules for the occurrence of the
sign-blessing in fermionic systems with momentum-imaginary-time
representation. 
One generic mechanism which is particularly important for fermions is
the crossing symmetry, as depicted in Fig. \ref{bases}, namely
permuting arbitrary two fermionic propagators causes an overall sign
change to the diagram.
If two fermionic propagators being exchanged carry similar momentum, which occurs near the Fermi surface, the direct and exchange diagrams strongly compensate with each other. It is therefore important to optimally arrange the internal variables so that the diagram integrand $W_{group}$ keeps the exact crossing symmetry.

Another generic mechanism is the conservation laws (or Ward identities). For example, the conserving diagrammatic expansions for the polarization proposed in the previous section satisfy $P({\bf{q}}\rightarrow 0, \tau)\rightarrow 0$ when approaching the zero temperature. However, this is an emergent property satisfied only by the sum of a conserving group of diagrams. In fact, all individual polarization diagrams (except the bubble diagram) break the conservation law and fluctuate around zero. Therefore, we observe a strong sign cancellation between the diagrams in the same conserving group. According to the Baym-Kadanoff approach in Eq. (\ref{eq:chi}), there is one-to-one correspondence between the minimal conserving groups for the polarization diagrams of the order $N$ and the $\ln Z$ diagrams of the order 
$N-1$
\footnote{To make the statement rigorous, all $\ln Z$ diagrams containing Hartree terms should be excluded from this correspondence since the polarization diagrams are one-interaction-irreducible.}. 
Indeed, for an arbitrary $\ln Z$ diagram, one can simply attach two external vertices to two of the bare electron propagators $g$ in all possible ways, and generate a conserving group for the polarization function. Strictly speaking, the sign blessing of the conserving groups is only guaranteed after integrating out all internal variables. However, provided that the internal variables of the polarization diagrams are inherited from the same free energy diagram, the operation of inserting two external vertices generates different time-ordered polarization diagrams, and leads to sign alternation within the conserving groups, implicitly encoding the sign blessing of the conservation law.

Now we are ready to propose the algorithm to group the diagrams and properly arrange internal variables. 
The algorithm is applicable to an arbitrary combination of momentum/frequency or space/time variables. 
To be consistent with the main text, we describe the algorithm with momentum/time representation. 
The main steps of the algorithm are:

i) Pick an arbitrary order-$N$ connected $\ln Z$ diagram, label all $2N$ time variables and choose $N+1$ \textit{independent} momentum loops. Keep momentum loops as short as possible.

ii) Generate a new connected $\ln Z$ diagram by permuting two electron propagators, rearrange the momentum loops as described in Fig.~\ref{bases} so that they automatically form a complete and independent loop basis for the new diagram. Thanks to the crossing symmetry, the new diagram has the opposite sign to the starting diagram. This step is repeated until all $\ln Z$ diagrams are exhausted.

iii) For each $\ln Z$ diagram, attach two external vertices to two of the electron propagators in all possible ways, to generate a conserving group of polarization diagrams. 
The arrangement of the internal variables of the original $\ln Z$ should not be modified in this step, so that the generated polarization diagrams share common parts of the diagram (many equal propagators).

It is also possible to apply the above algorithm to Hugenholtz diagrams, which form a particular subset generated by the algorithm in Fig.~(\ref{bases}) (when the top and the bottom bosonic propagators are connected to each other). These diagrams combine the direct and exchange interaction into an antisymmetric four-point vertex. 
They are particularly convenient if one works with momentum/frequency, or momentum/time representation, and the interaction is instantaneous, as in our model Eq. (\ref{eq:L0}) and Eq. (\ref{eq:dL}).

Finally, we briefly discuss the benefits of grouping
  the diagrams in the diagrammatic Monte Carlo
  algorithm. There are two improvements in terms of the Monte Carlo
  efficiency. First, the total weight function $F_2^N$ sampled by the
  Markov chain is much smaller than $F_3^N$. This indicates the
  variance of the integrand is dramatically reduced, which improves
  the statistical error. Second, the diagrams in the same group
  typically share many common objects (propagators and
  interactions). This simplifies the total diagram weight calculations
  in each Monte Carlo update. For example, all Feynman diagrams (up to
  $2^N$ diagrams at order $N$) that belong to the same Hugenholtz
  diagram, share the same set of propagators, thus they only need to
  be evaluated once. Indeed, all Feynman diagrams that belong to the
  same Hugelholtz diagram can be choosen to have all fermionic
  propagators identical. Those are computed once, and not $2^N$
  times. 
Furthermore, the interaction lines are not identical, however, they
contain a lot of common products. One can show that a binary tree can
be constructed, with the depth equal to the number of Hugenholtz
interaction propagators, in which each vertex of the binary tree adds
either the direct or the exchange interaction to the Hugenholtz
diagram. 
The leaves of such a binary tree contain exactly $2^N$ terms,
corresponding to the products we need to evaluate the sum of $2^N$
Feynman diagrams, 
while the number of operations to evaluate such a tree grows as $O(N)$.

\end{document}